\documentclass[journal=jacsat,manuscript=article]{achemso}
\usepackage[version=3]{mhchem} 
\usepackage{ulem}
\usepackage{amsmath}
\usepackage{amssymb}
\usepackage{mathtools,cases}
\usepackage{stfloats}
\usepackage{xcolor}
\usepackage{float}
\usepackage{float}
\usepackage{indentfirst}
\usepackage{comment}

\usepackage{lineno} 

\author{Volodymyr Vasylkovskyi}
\affiliation[JMU]{Experimental Physics 6 and Würzburg-Dresden Cluster of Excellence ct.qmat, Julius-Maximilian University of Würzburg, 97074 Würzburg, Germany}
\email{volodymyr.vasylkovskyi@uni-wuerzburg.de}
\author{Andrey B. Evlyukhin}
\affiliation[IQO]{Institute of Quantum Optics and Cluster of Excellence PhoenixD, Leibniz University Hannover, 30167 Hannover, Germany}
\email{evlyukhin@iqo.uni-hannover.de}
\author{Elena Schlein}
\affiliation[IQO]{Institute of Quantum Optics and Cluster of Excellence PhoenixD, Leibniz University Hannover, 30167 Hannover, Germany}
\alsoaffiliation[KhPI] {National Technical University "Kharkiv Polytechnic Institute", 61002 Kharkiv, Ukraine}
\author{Mykola Slipchenko}
\affiliation[ISMA]{Institute for Scintillation Materials, National Academy of Sciences of Ukraine, 61072 Kharkiv, Ukraine}
\alsoaffiliation[KhPI] {National Technical University "Kharkiv Polytechnic Institute", 61002 Kharkiv, Ukraine}
\author{Roman Kiyan}
\affiliation[IQO]{Institute of Quantum Optics and Cluster of Excellence PhoenixD, Leibniz University Hannover, 30167 Hannover, Germany}
\author{Kestutis Kurselis}
\affiliation[IQO]{Institute of Quantum Optics and Cluster of Excellence PhoenixD, Leibniz University Hannover, 30167 Hannover, Germany}
\author{Vladimir Dyakonov}
\affiliation[JMU]{Experimental Physics 6 and Würzburg-Dresden Cluster of Excellence ct.qmat, Julius-Maximilian University of Würzburg, 97074 Würzburg, Germany}
\author{Boris Chichkov}
\affiliation[IQO]{Institute of Quantum Optics and Cluster of Excellence PhoenixD, Leibniz University Hannover, 30167 Hannover, Germany}

\title
  { Synthesis of organic-inorganic perovskite and all-inorganic lead-free double perovskite nanocrystals by femtosecond laser pulses }

\abbreviations{IR,NMR,UV}
\keywords{American Chemical Society, \LaTeX}

\begin{document}








\begin{abstract}

{Perovskite materials are at the forefront of modern materials science due to their exceptional structural, electronic, and optical properties. The controlled fabrication of perovskite nanostructures is crucial for enhancing their performance, stability, and scalability, directly impacting their applications in next-generation devices such as solar cells, LEDs, and sensors. Here, we present a novel, ligand-free approach to synthesize perovskite nanocrystals (NCs) with average sizes up to 100 nm, using femtosecond pulsed laser ablation (PLA) in ambient air without additional liquid media. We demonstrate this method for both organic-inorganic  (methylamino lead)  hybrid perovskites (MAPbX$_3$, X = Cl, Br, I) and fully inorganic lead-free double perovskites (Cs$_2$AgBiX$_6$, X = Cl, Br), achieving high-purity NCs without stabilizing ligands — a critical advancement over conventional chemical synthesis methods. By tailoring laser parameters, we systematically elucidate the influence of perovskite composition (halide type, organic vs. inorganic cation, single vs. double perovskite structure) on the ablation process and the resulting nanocrystal properties. Transmission electron microscopy and X-ray diffraction confirm the preservation of crystallinity, with MAPbX$_3$ forming larger ($\sim 90$ ~nm) cubic NCs and Cs$_2$AgBiX$_6$ forming smaller ($\sim 10$~nm) rounded NCs. Photoluminescence spectroscopy reveals pronounced size-dependent blue shifts (17–40 nm) due to quantum confinement, particularly for Br- and I-containing perovskites.
This clean, scalable, and versatile PLA approach not only provides direct access to high-purity, ligand-free perovskite NCs with tunable optical properties but also represents a significant advance in the fabrication of nanostructures, enabling the exploration of new perovskite-based optoelectronic and quantum devices.}
\end{abstract}

\section{Introduction}

Pulsed laser ablation (PLA) is a versatile top-down method widely employed for the synthesis of nanomaterials, including metals, oxides, conventional semiconductors, and, more recently, complex compounds such as perovskites \cite{zhang2017colloidal,zywietz2014generation,vasylkovskyi2023laser,rosa2019laser,tselikov2025tunable,Vasylkovskyi_2023}. The technique relies on a laser system with optical focusing, which may include positioning or scanning devices, along with a solid target. Short laser pulses are focused on the target surface, locally concentrating energy and inducing ablation, leading to the formation of nanoparticles (NPs). The main advantages of PLA are its versatility and the high purity of the products compared to alternative methods. The efficiency of the process critically depends on precise control of laser parameters such as wavelength, fluence, pulse duration, and repetition rate \cite{du2020rare}.

In contrast to bottom-up chemical methods (ligand-assisted reprecipitation, hot-injection, sol-gel, solvothermal synthesis, etc.), which typically require organic ligands (e.g., oleic acid (OA) or oleylamine (OAm)) to stabilize particles and prevent aggregation, PLA enables the production of nanocrystals (NCs) without organic capping agents. These ligands can hinder charge transport across the surface of NCs and limit their applications \cite{Biswas_2023}. Moreover, ligand stabilization is prone to degradation under UV irradiation or prolonged storage in liquid media \cite{vasylkovskyi2023electrochemiluminescence}. Laser-based ligand-free approach is particularly advantageous for electrochemical sensing and catalytic applications, where surface chemistry, efficient charge transport, and minimal post-synthesis processing are crucial. In contrast to other top-down techniques such as ball milling - which is limitated by the minimal achievable nanoparticle size and possibility to introduce impurities - PLA enables fabrication of nanoparticles below 100 nm, while ensuring high phase and surface purity. Thus, this method offers a unique and scalable route for the synthesis of high-purity halide NCs, opening opportunities for the development of functional nanomaterials that cannot be easily obtained by conventional methods.

Recent reviews \cite{berestennikov2019active,cherepakhin2024advanced} highlight the significance of laser processing and nanostructuring of perovskites for advanced light-emitting systems and nanoscale light management. Laser techniques, including laser nucleation, annealing, printing, and patterning \cite{zhang2025application}, provide unprecedented control over material properties. Among these, femtosecond PLA is especially noteworthy due to its high precision and minimal thermal impact. 

Despite previous studies, systematic investigations into the laser-assisted synthesis of halide perovskite NCs, particularly the comparison between hybrid and fully inorganic lead-free perovskites, remain extremely limited. Halide perovskites, unlike more robust crystalline compounds (chalcogenides, classical semiconductors, oxide perovskites), require careful optimization of PLA conditions to prevent thermal decomposition. Multicomponent crystals are often sensitive to thermal stress, which can induce cracking, phase transitions, or degradation. Overheating can be mitigated by using a liquid medium in the ablation zone or employing femtosecond pulses, which minimize thermal diffusion and preserve crystallinity under ambient conditions. This approach also avoids solvent-induced transformations previously observed in toluene and chloroform. In our previous studies \cite{vasylkovskyi2023laser}, PLA of CeAlO$_3$ single crystals enabled the synthesis of NCs that retained crystallinity and exhibited enhanced optical properties, demonstrating the practical potential of PLA for creating nanomaterials with superior functional properties. Further development of PLA for the generation of new nanomaterials is one of the most important prospect of modern photonics.

Halide perovskites with the general ABX$_3$ structure (A is the monovalent cation, B is the metal cation, X is the halide: Cl\(^-\), Br\(^-\), I\(^-\)) are particularly attractive due to their compositional flexibility and wide range of applications. Hybrid organic-inorganic compounds such as methylammonium lead halide perovskite, MAPbX$_3$ and formamidinium lead halide perovskite, FAPbX$_3$ display structural features that influence exciton confinement, ion migration, and lattice stability. At the nanoscale, they show enhanced luminescence, size-dependent optical properties, and high surface reactivity, making them promising candidates for light-emitting diodes, lasers, and catalysis. The doped compositions of perovskite NCs may also be promising for spintronic applications.

Additionally, lead-free double perovskites A$_2$B$^+$B$^{3+}$X$_6$ (e.g., Cs$_2$AgBiCl$_6$, Cs$_2$AgBiBr$_6$) are actively studied, where two metal cations substitute Pb$^{2+}$ while maintaining charge balance \cite{zhao2018rational}. These compounds are highly stable under ambient conditions and are considered promising materials for scintillation detectors, photodetectors, and quantum technologies. To our knowledge, there are no prior reports demonstrating a direct, systematic comparison of ligand-free NCs derived from both hybrid MAPbX$_3$ and fully inorganic Cs$_2$AgBiX$_6$ using femtosecond PLA. This gap highlights the novelty and scientific importance of the present work.

In this study, we demonstrate the first application of femtosecond PLA in ambient air without additional liquid media as a green, scalable, and universal method for synthesizing ligand-free halide NCs across two structurally distinct perovskite classes.  Hybrid organic-inorganic MAPbX$_3$ (X = Cl, Br, I) and fully inorganic lead-free Cs$_2$AgBiX$_6$ (X = Cl, Br) NCs were synthesized and systematically characterized. Their structural and optical properties were analyzed using transmission electron microscopy (TEM), X-ray diffraction (XRD), UV–visible absorption, and photoluminescence (PL). This work provides critical insights into the relationship between perovskite composition and laser ablation dynamics, offering a novel platform for the rational design of high-performance perovskite nanomaterials with potential applications in optoelectronics, quantum technologies, and nanophotonics.

\section{Materials and methods}

\subsection{Crystal growth}

Organic-inorganic perovskite single crystals, as  targets for laser ablation,  were grown by the inverse temperature crystallization (ITC) technique \cite{Saidaminov_2015}. Materials used for crystal growth included: methylammonium chloride (MACl, CH$_3$NH$_3$Cl, 99.5\%), methylammonium bromide (MABr, CH$_3$NH$_3$Br, 99.5\%), methylammonium iodide (MAI, CH$_3$NH$_3$I, 99.5\%) and lead(II) chloride (PbCl$_2$, 99.999\%) from “Lumtec”; lead(II) bromide (PbBr$_2$, 99\%) and lead(II) iodide (PbI$_2$, 99.99 \%) from TCI. A 1:1 (v/v) mixture of dimethylformamide (DMF, C$_3$H$_7$NO, 99.8\%) and dimethyl sulfoxide (DMSO, C$_2$H$_6$OS, 99.99\%) from “Sigma-Aldrich” was used as both solvent and crystal growth medium. All reagents were used without further purification. To prevent degradation, the reagents were handled and weighed in a nitrogen-filled glovebox with a water and oxygen content of less than 2 ppm. The precursor solutions of MAPbCl$_3$, MAPbBr$_3$, MAPbI$_3$ were prepared and placed in clean glass vials, which were sealed and positioned in a silicon oil bath on the hot plate. The temperature gradually increased at a rate of 1\:$^{\rm o}$C per hour until 60\:$^{\rm o}$C for MAPbCl$_3$, MAPbBr$_3$ and 110\:$^{\rm o}$C for MAPbI$_3$ were reached. The resulting single crystals were collected, washed with dichloromethane to remove surface residues, and dried under ambient conditions.

All-inorganic lead-free double perovskite crystals,  as other targets for laser ablation, were grown using a controlled cooling crystallization technique \cite{Armer_2021}. The following precursors were used: cesium chloride (CsCl, 99.9\%), cesium bromide (CsBr, 99.999\%), silver chloride (AgCl, 99,999\%), and bismuth trichloride (BiCl$_3$, 98\%) from “Sigma-Aldrich”; Silver bromide (AgBr, 99\%) and bismuth tribromide (BiBr$_3$, 98\%) from “Alfa”. Hydrochloric acid (HCl, 36\%) from “ThermoScientific” and hydrobromic acid (HBr, 48\%) from “Acros Organics” were used as solvents and crystal growth media for Cs$_2$AgBiCl$_6$ and Cs$_2$AgBiBr$_6$, respectively. The flask containing the chemicals and solvent was placed in a silicon oil bath on the hot plate. The oil bath was heated to 160 °C for Cs$_2$AgBiCl$_6$ and 120$^{\rm o}$C for Cs$_2$AgBiBr$_6$ and maintained at the set temperature for 5 hours to ensure complete dissolution. The temperature then slowly decreased at the rate of 1$^{\rm o}$C per hour to room temperature to allow controlled nucleation and crystal growth. The resulting single crystals were harvested, washed with dichloromethane, and dried under ambient conditions. Cs$_2$AgBiI$_6$ was excluded from the research due to its thermodynamic instability.

\subsection{Pulsed laser ablation}

Perovskite NCs were fabricated via PLA in ambient conditions using a femtosecond laser system - CARBIDE with a harmonics module from Light Conversion UAB. The laser parameters were as follows: wavelength 343 nm, pulse duration 250 fs, repetition rate 100 kHz. The laser beam was focused onto the surface of selected single-crystal perovskites, enclosed in a glass vial, using a 50 mm focal distance aplanatic doublet lens (Fig. S1 in Supporting information (SI)).

PLA was performed by scanning laser pulses over the crystal surface using a combined motion of large-scale translation XYZ stage (PI miCos GmbH) and fast wobbling by a scanner (excelliSCAN 14, SCANLAB GmbH). The resulting velocity of the laser beam spot on the sample surface was 0.12 m/s. PLA was performed under ambient air without the use of additional liquid or gas media.

\subsection{Characterization of Perovskite single crystals and nanocrystals}

Optical characterization included UV-Vis absorption spectroscopy, performed using a Jasco V-630 spectrometer. Steady-state photoluminescence (PL) spectra were recorded using an Edinburgh Instruments FLS 980 spectrometer, with excitation at 375 nm. 

Transmission Electron Microscopy (TEM) measurements were performed on a Tecnai G2 F20 TMP instrument (FEI). To perform the measurements, NCs were placed in chloroform and drop-casted onto carbon-coated copper grids and dried under ambient conditions.

X-Ray Diffraction (XRD)  measurements were performed using a General Electric XRD 3003 TT system with a monochromatic copper (Cu) K$\alpha$ radiation source to determine the crystalline structure. For powder XRD analysis, single crystals were manually ground into fine powders, while PLA-fabricated NCs were directly collected as nanopowders; both were then uniformly spread onto a polymer substrate. Measurements were performed under ambient conditions in Bragg-Brentano geometry within the optimal 2Theta range with a step size of 0.005 ° and a dwell time of 5 seconds per step.

\section{Results and discussion}

PLA was carried out in ambient air without additional liquid media to reduce the interference of media and fabricated NCs with the laser beam (Fig. S1, S2 in Supporting information (SI)). Solvent-induced transformations, such as those observed with toluene or chloroform \cite{vasylkovskyi2023laser}, were avoided, ensuring the fabrication of high-purity NCs without contamination and enabling straightforward characterization of the resulting NCs. PLA led to the successful synthesis of perovskite NCs, which exhibited notable distinctions in optical and structural properties compared to their bulk counterparts.

\subsection{Pulsed laser ablation of organic-inorganic halide perovskites}

PLA of organic-inorganic halide perovskite bulk crystals led to the fabrication of nanodispersed powders of MAPbCl$_3$, MAPbBr$_3$, and MAPbI$_3$. The laser pulse energy was optimized for each perovskite to achieve the highest ablation throughput, which, under the investigated experimental conditions, was also influenced by the limited and confined process space: 1.7 µJ for MAPbCl$_3$, 0.76 µJ for MAPbBr$_3$, and 0.53 µJ for MAPbI$_3$. That is in contradiction to the structural properties of these materials (of which MAPbCl$_3$ is the most unstable kinetically and thermodynamically compared to MAPbBr$_3$ and MAPbI$_3$, which are almost identically stable) and could rather indicate either the relation to the corresponding B-X bond lengths of these crystals (e.g. Pb-Cl: 2.857 Å; Pb-Br: 2.973 Å; Pb-I: 3.164 Å) or to the variation in absorption of laser irradiation \cite{wiedemann2021hybrid,dang2015bulk,islam2019mixed,brunetti2016thermal}.

The color of the resulting nanopowders was generally consistent with that of the corresponding bulk crystals. MAPbCl$_3$ NCs appeared white (Fig. S3b1 in SI) compared to the transparent crystal (Fig. S3a1 in SI), and MAPbBr$_3$ NCs appeared to be yellow (Fig. S3b2 in SI) compared to the orange bulk single crystal (Fig. S3a2 in SI), probably due to the quantum confinement effect. Compared to black single crystals (Fig. S3a3 in SI), MAPbI$_3$ NCs under ambient conditions appeared to be dark yellow (Fig. S3b3 in SI), due to the partial degradation of the MA organic molecules under exposure to moisture in the air, with the following appearance of yellow PbI$_2$ on the surface of NCs. However, MAPbI$_3$ NCs in colloidal form in chloroform appear black, corresponding to the bulk crystal (Fig. S3c3 in SI). Therefore, organic-inorganic perovskite NCs synthesized in ambient conditions should nevertheless be stored in nonpolar solvents to prevent moisture-induced degradation.

Photoluminescence spectra revealed sharp emission peaks (full width at half maximum (FWHM): 17 – 61 nm) for NCs and bulk crystals (Fig. 1a). The MAPbCl$_3$ bulk crystal exhibited two PL peaks at 424 nm, which could be related to NCs, and 434 nm (Fig. 1a1), which can be attributed to crystals bigger than 300 nm that were evidenced by TEM measurements (Fig. S4 in SI). Pronounced blue shifts of the PL peaks of NCs towards higher energies compared to the bulk crystals were observed for MAPbBr$_3$ (from 573 to 530 nm) (Fig. 1a2) and for MAPbI$_3$ (from 803 to 762 nm) (Fig. 1a3), with shifts of $\sim 40$ nm (0.18 eV) attributed to the quantum confinement effect. The transition from bulk to nanoscale increases bandgap energy by confining electrons and holes within the smaller volume of NCs, resulting in shorter-wavelength PL emissions compared to bulk, and such a shift could be limited with smaller NCs' size \cite{Wen_2015}.

UV-Vis absorption measurements of the colloidal solutions showed sharp absorption peaks at 404 nm for MAPbCl$_3$ (Fig. 1b1), 527 nm for MAPbBr$_3$ (Fig. 1b2), and a broad peak at 537 nm for MAPbI$_3$ (Fig. 1b3), which correspond to the data reported in the literature \cite{Matsushima_2019, Khan_2023, Chen_2019}.

\begin{figure}[H]
    \centering
    \includegraphics[width=1\linewidth]{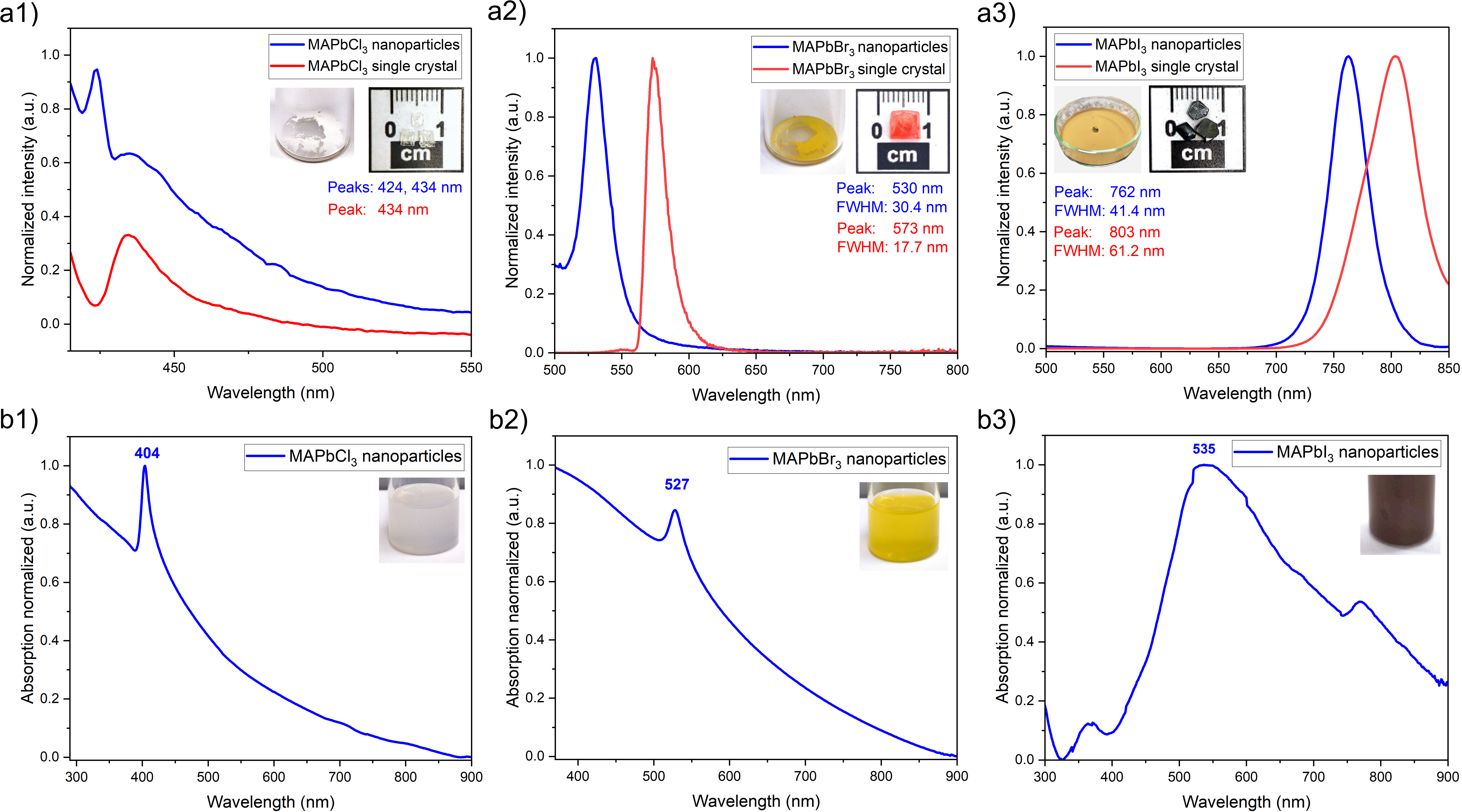}    
    \caption{(a) Photoluminescence spectra of (a1) MAPbCl$_3$, (a2) MAPbBr$_3$, (a3) MAPbI$_3$ nanocrystals and bulk crystals. Insets: optical images of single crystals and nanopowders. Insets: images of nanopowder and bulk single crystals; (b) UV-Vis absorption spectra of (b1) MAPbCl$_3$, (b2) MAPbBr$_3$, and (b3) MAPbI$_3$ NCs in colloid state. Insets: colloidal solutions in chloroform.
}
    \label{fig_1}
\end{figure}

TEM measurements (Fig. 2a) revealed the successful fabrication of NCs of organic-inorganic perovskites. Although the morphology was somewhat inhomogeneous, a substantial number of cubic-shaped NCs were observed, particularly for MAPbCl$_3$ (Fig. 2a1) and MAPbBr$_3$ (Fig. 2a2). The cubic shape is usually a characteristic of NCs synthesized via bottom-up chemical routes rather than top-down approaches \cite{HUANG2020116984}. The cubic morphology of the NCs reflects the cubic phase of the perovskite, likely resulting from laser-induced cleavage along the crystal planes during the laser ablation. MAPbI$_3$ NCs had more non-uniform shapes with corners and flat edges; however, cubic NCs were also noticeable (Fig. 2a3). The appearance of voids in MAPbBr$_3$ NCs was noticed, which may result from laser-induced thermal stress or MA degradation (Fig. 2a2). The presence of atomic planes and fast Fourier transform (FFT) patterns in high-resolution images confirms that the crystallinity of laser-fabricated nanomaterials has been preserved (insets in Fig. 2a). The size distribution of NCs was calculated from multiple TEM images of each perovskite (Fig. S4, S5, S6 in SI) and appeared normal, with mean sizes of 107.31 nm for MAPbCl$_3$ (Fig. 2b1), 59.88 nm for MAPbBr$_3$ (Fig. 2b2), and 66.8 nm for MAPbI$_3$ (Fig. 2b3). However, the prolonged agglomerates consisted of MAPbI$_3$ NCs with 3 - 20 nm sizes were noticed (Fig. S7 in SI). The standard deviations of 71.88 nm, 39.76 nm, and 43.14 nm, respectively, indicate varying degrees of size uniformity, with the higher deviation for MAPbCl$_3$ suggesting greater size variability that may influence optical properties such as photoluminescence spectra.

\begin{figure}[H]
    \centering
    \includegraphics[width=1\linewidth]{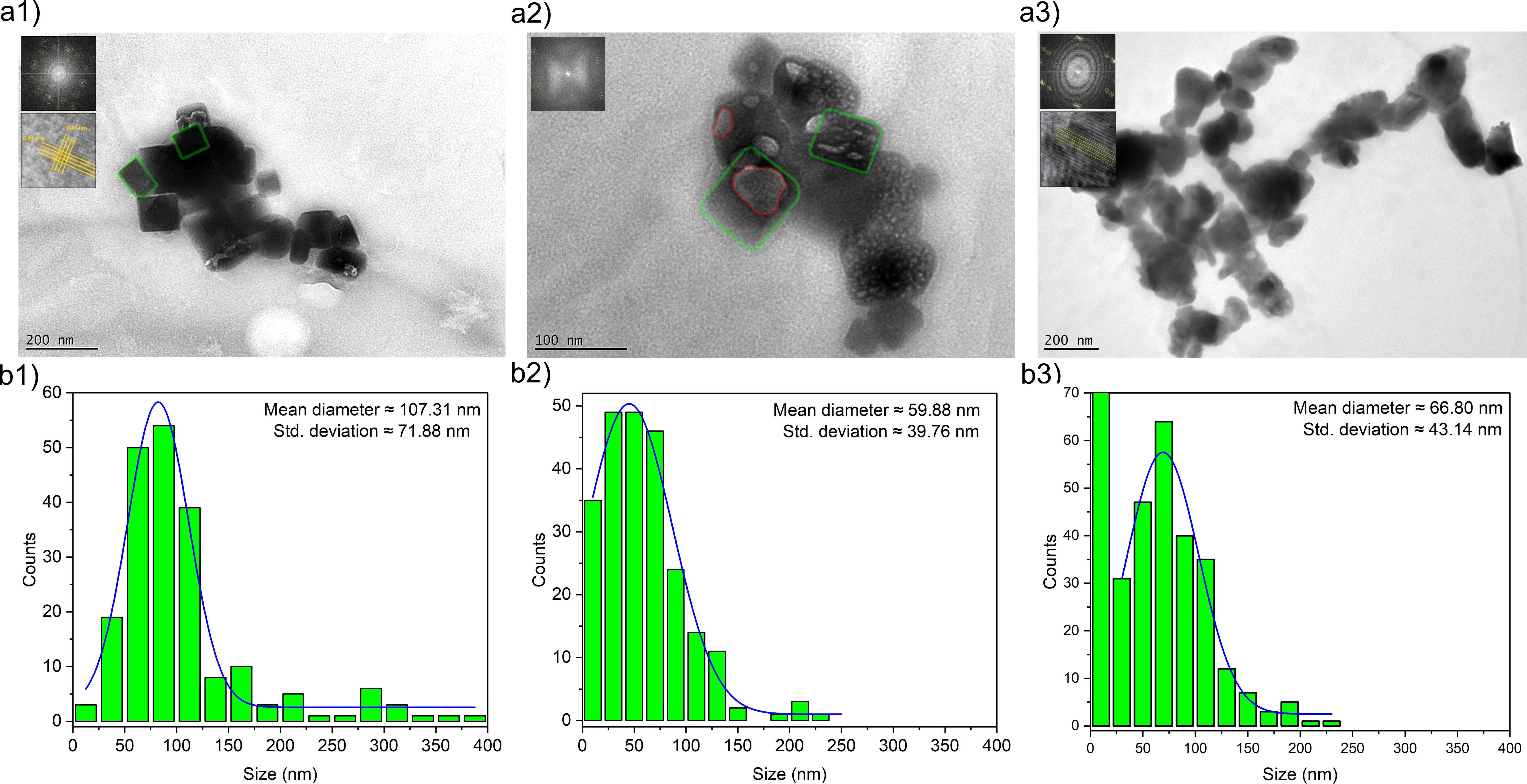}    
    \caption{(a) TEM images of (a1) MAPbCl$_3$, (a2) MAPbBr$_3$, and (a3) MAPbI$_3$ NCs (Top-left inserts: corresponding fast Fourier transform and lattice planes); (b) Size distribution, mean diameter, and standard deviation of (b1) MAPbCl$_3$, (b2) MAPbBr$_3$, and (b3) MAPbI$_3$ NCs}
    \label{fig_2}
\end{figure}

XRD patterns of MAPbCl$_3$ (Fig. 3a), MAPbBr$_3$ (Fig. 3b), and MAPbI$_3$ (Fig. 3c) NCs, along with powder XRD patterns of grown single crystals, were measured to confirm phase purity in all crystal planes. The peak positions of the XRD patterns of laser-fabricated NCs and the powder XRD of the crystals align with each other and correspond to simulated patterns derived from confirmed crystallographic data \cite{islam2019mixed,dang2015bulk,wiedemann2021hybrid}, with no additional peaks observed, which confirms the absence of additional crystal phases. As expected for nanoscale materials, the NCs’ diffraction patterns exhibit peak broadening, along with reduced signal intensity and higher noise compared to the powder XRD pattern due to: smaller crystallite sizes, which reduce the number of diffracting planes; increased surface disorders from a high surface-to-volume ratio; and possible strain.

\begin{figure}[H]
    \centering
    \includegraphics[width=1\linewidth]{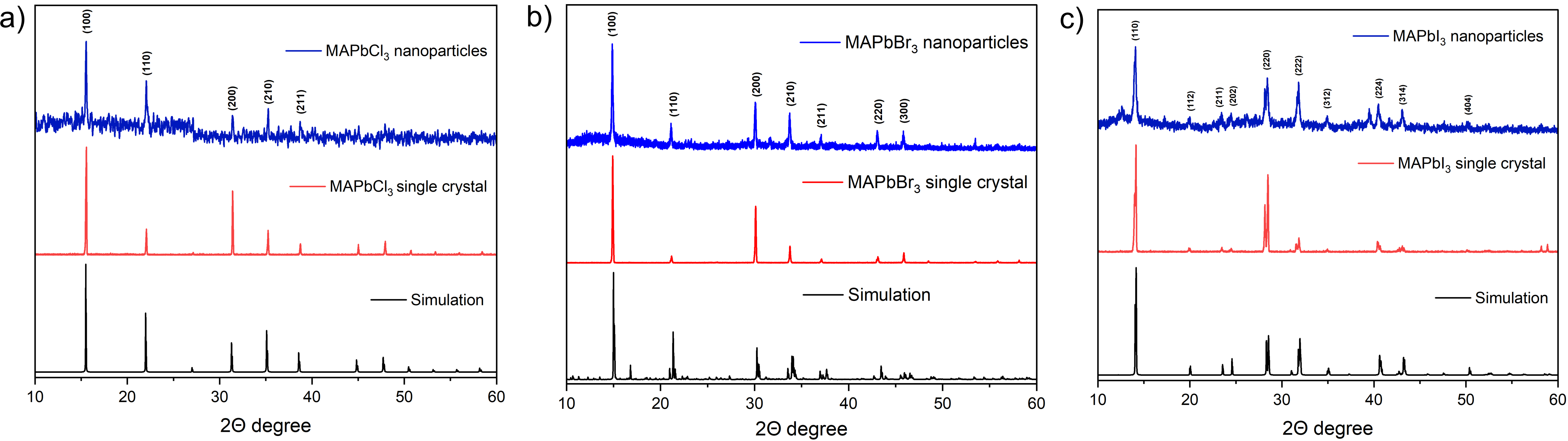}    
    \caption{XRD patterns of (a) MAPbCl$_3$, (b) MAPbBr$_3$, and (c) MAPbI$_3$ NCs compared to powder XRD of bulk crystals and XRD patterns simulated from CIF data of \cite{islam2019mixed,dang2015bulk,wiedemann2021hybrid}
}
    \label{fig_3}
\end{figure}

\subsection{Pulsed laser ablation of all-inorganic lead-free double perovskites }

PLA of all-inorganic lead-free double perovskites revealed similar trends as in the case of hybrid perovskites. Comparable amounts of Cs$_2$AgBiCl$_6$ and Cs$_2$AgBiBr$_6$ nanopowders were produced using the same laser parameters except the laser pulse energy, which was lowered to 0.76 µJ in order to achieve high ablation throughput comparable to MAPbX$_3$ perovskites. The shorter B-X and B'-X bond lengths in Cs$_2$AgBiX$_6$ (Ag-Cl: 2.708 Å; Bi-Cl: 2.681 Å for Cs$_2$AgBiCl$_6$, and Ag-Br: 2.821 Å; Bi-Br: 2.814 Å for Cs$_2$AgBiBr$_6$) suggest higher lattice stability compared to MAPbX$_3$, yet broader PL peaks may indicate increased defect formation due to higher defect tolerance of the crystal lattice because of B/B’ cation complexity \cite{jain2013commentary,slavney2016bismuth} (Fig. 4a). Unlike organic-inorganic perovskite NCs, all-inorganic double perovskite NCs demonstrate enhanced stability under ambient conditions, due to the presence of Cs instead of more degradable organic MA component.

The color of Cs$_2$AgBiCl$_6$ NCs appeared white (Fig. S8b1 in SI) comparing to the corresponding yellow bulk crystal (Fig. S8a1 in SI), while Cs$_2$AgBiBr$_6$ powder exhibited a yellow color (Fig. S8b2 in SI) in contrast to the red color of the bulk crystal (Fig. S8a2 in SI).  PL spectra of NCs revealed broad emission peaks for both NCs and bulk crystals of Cs$_2$AgBiX$_6$ evidenced by 4 times larger FWHM (125 – 240 nm) than for hybrid perovskites (Fig. 4a). The broad PL peaks of both double perovskites may be attributed not solely to defect states, but also to electron-phonon coupling or self-trapped excitons \cite{Zhu_2023,Luo_2018}. A blue spectral shift of 30 nm (0.09 eV) was observed in Cs$_2$AgBiBr$_6$ compared to the bulk crystal, indicating a quantum confinement effect. In contrast, Cs$_2$AgBiCl$_6$, with similar average size, exhibited a smaller red shift of 17 nm (0.052 eV), with overlapping PL peaks at 645 nm and 678 nm. UV-Vis absorption spectra of colloidal NCs dispersions showed absorption peaks at 379 nm for Cs$_2$AgBiCl$_6$ (Fig. 4b1), and 485 nm for Cs$_2$AgBiBr$_6$ (Fig. 4b2). 

\begin{figure}[H]
    \centering
    \includegraphics[width=0.8\linewidth]{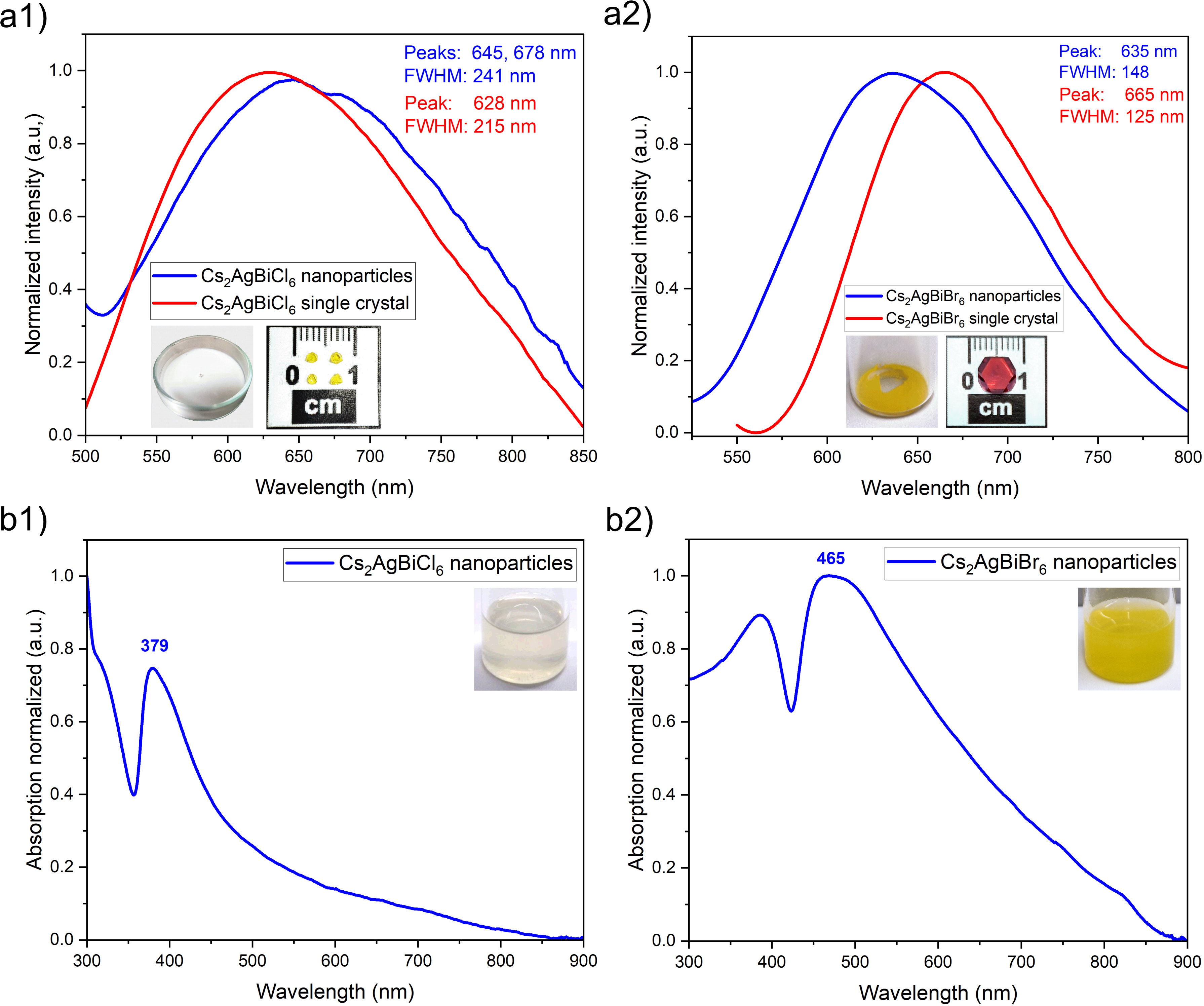}    
    \caption{(a) Photoluminescence spectra of (a1) Cs$_2$AgBiCl$_6$ and (a2) Cs$_2$AgBiBr$_6$ NCs and bulk crystals. Insets: optical images of single crystals and nanopowders. Insets: images of nanopowder and bulk single crystals; (b) UV-Vis absorption spectra of (b1) Cs$_2$AgBiCl$_6$ and (b2) Cs$_2$AgBiBr$_6$ NCs in colloid state. Insets: colloidal solutions in chloroform.
}
    \label{fig_4}
\end{figure}

The TEM measurements confirmed the formation of NCs with rounded shape, with less frequent appearance of flat edges and corners compared to MAPbX$_3$ NCs (Fig. 5a). The crystallinity was evidenced by the presence of crystal planes on TEM images as well as FFT patterns (insets in Fig. 5a). The morphology of laser-fabricated Cs$_2$AgBiX$_6$ NCs was assessed using multiple TEM images (Fig. S9, S10 in SI), revealing a normal size-distribution with the average size of 11.37 nm for Cs$_2$AgBiCl$_6$ (Fig. 5b1) and 15.65 nm for Cs$_2$AgBiBr$_6$ (Fig. 5b2). High standard deviations of 5.05 nm for Cs$_2$AgBiCl$_6$ and 7.54 nm for Cs$_2$AgBiBr$_6$ indicate a significant diversity in NC sizes, comparable to MAPbX$_3$ NCs. The smaller size and spherical morphology of Cs$_2$AgBiX$_6$ NCs compared to MAPbX$_3$ may result from a higher fragmentation rate, due to the defect-prone structure of double perovskites. 

\begin{figure}[H]
    \centering
    \includegraphics[width=0.8\linewidth]{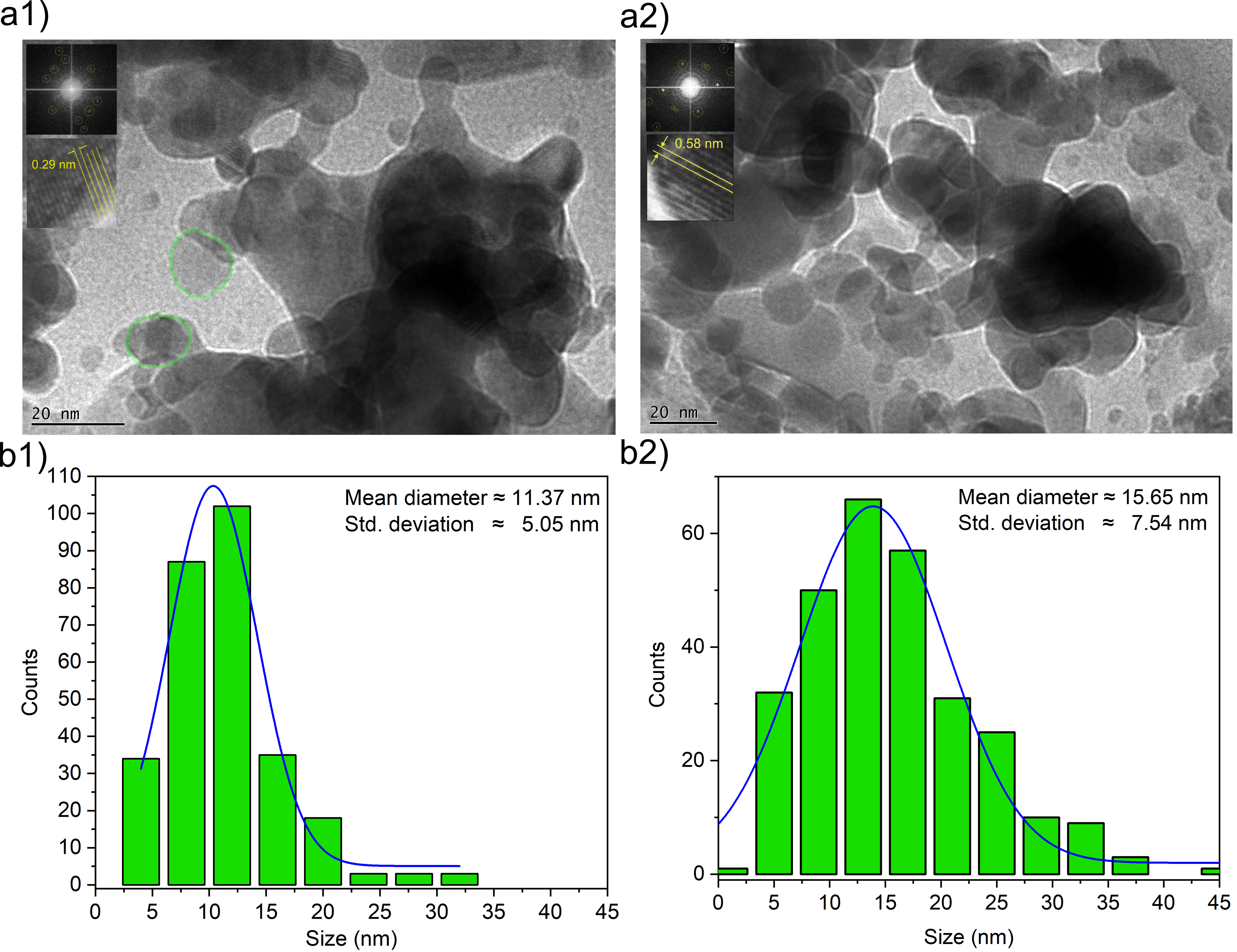}    
    \caption{(a) TEM images of (a1) Cs$_2$AgBiCl$_6$ and (a2) Cs$_2$AgBiBr$_6$ NCs. Top-left inserts: corresponding fast Fourier transform and lattice planes; (b) Size distribution, mean diameter, and standard deviation of (b1) Cs$_2$AgBiCl$_6$ and (b2) Cs$_2$AgBiBr$_6$ NCs.
}
    \label{fig_5}
\end{figure}

XRD patterns of the laser-fabricated Cs$_2$AgBiCl$_6$ (Fig. 6a)  and Cs$_2$AgBiBr$_6$ (Fig. 6b) NCs are in good agreement with those of the powder XRD patterns of bulk crystals and simulated patterns derived from confirmed crystallographic data \cite{McClure_2016,slavney2016bismuth}. No additional peaks or peak splitting were observed, indicating that the phase purity and structural integrity of perovskites were preserved after ablation. The increased noise ratio for the XRD pattern from Cs$_2$AgBiBr$_6$ is attributed to the smaller amount of the material used for the measurement.

\begin{figure}[H]
    \centering
    \includegraphics[width=0.8\linewidth]{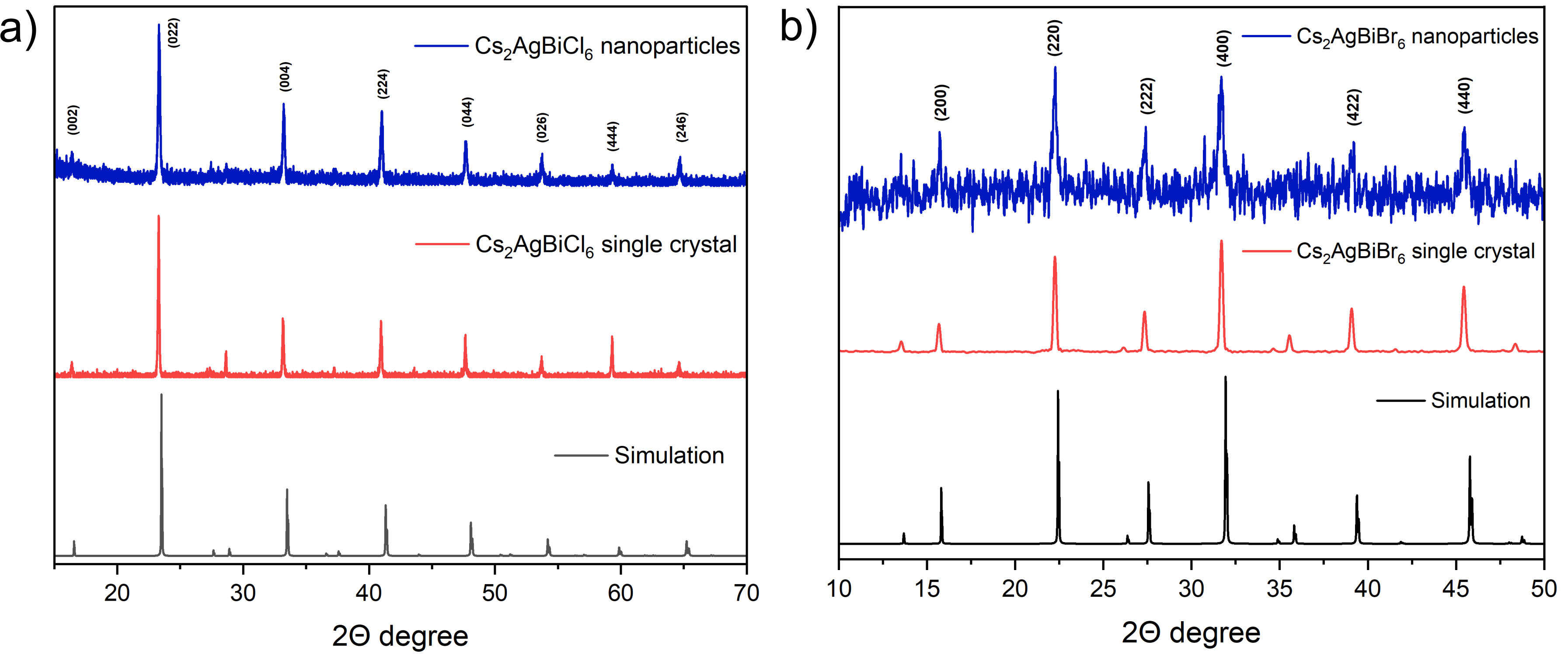}
    \caption{XRD patterns of (a) Cs$_2$AgBiCl$_6$, and (b) Cs$_2$AgBiBr$_6$ NCs compared to powder XRD of bulk crystals and XRD patterns simulated from CIF data of  \cite{McClure_2016,slavney2016bismuth}}
    \label{fig_6}
\end{figure}

\section{Conclusions}

Femtosecond pulsed laser ablation in ambient air has been successfully applied to synthesize organic-inorganic MAPbX$_3$ perovskite (X = Cl, Br, I) and all-inorganic lead-free double perovskite Cs$_2$AgBiX$_6$ (X = Cl, Br) NCs, enabling the production of ligand-free NCs without the use of additional synthesis media. TEM measurements confirmed the formation of MAPbX$_3$ NCs with mean sizes up to 107 nm, while Cs$_2$AgBiX$_6$ were smaller with 11-15 nm mean sizes, indicating a higher fragmentation ratio of double perovskites. It has been evidenced that MAPbX$_3$ NCs have a cubic shape, which may be related to laser-induced cleavage along the crystal planes of the cubic perovskite phase, whereas Cs$_2$AgBiX$_6$ NCs have a spherical shape without flat edges and corners. XRD confirmed the preserved crystallinity of laser-fabricated NCs, with no secondary phases detected. Photoluminescence studies revealed size-dependent spectral shifts (17-40 nm) attributed to quantum confinement, with more significant shifts in MAPbX$_3$ (X = Br, I) perovskites, highlighting the potential for optical property tuning via size manipulation. The broader PL emission observed in Cs$_2$AgBiX$_6$ NCs is consistent with their indirect band gaps and may arise from defect-related recombination typical of defect-tolerant double perovskite structure. 

The ligand-free surfaces of laser-fabricated NCs improve surface tunability and may enhance charge transport efficiency compared to chemically synthesized NCs, where capping ligands act as insulating barriers to charge transfer. These advances are beneficial for further electrochemical, biomedical, and photocatalytic applications.  This work positions PLA as a scalable, eco-friendly technique for producing high-purity perovskite NCs, extending their versatility and paving the way for novel perovskite-based devices.

\begin{acknowledgement}

We acknowledge financial support from the DFG, German Research Foundation under Germany’s Excellence Strategy within the Cluster of Excellence ct.qmat (EXC 2147, Project ID 39085490) and Cluster of Excellence PhoenixD (EXC 2122, Project ID 390833453). 
Thanks go to the Laboratory of Nano and Quantum Engineering (LNQE) of Leibniz University Hannover for the possibility of conducting TEM measurements.
\end{acknowledgement}

\section{Declarations}

Conflict of interest - The authors declare that they have no conflict of interest.

\section{Data Availability}

The data obtained and/or analyzed during the study are not publicly available but are available from the corresponding author upon reasonable request.

\bibliography{Perovskite}

\end{document}